\documentclass{iaus}
\usepackage{graphicx}
\usepackage{amsmath,epsfig,rotating,amssymb}
\usepackage{natbib}

\newcommand{\hii}{H~{\sc ii}}
\newcommand{\uchii}{UC~H~{\sc ii}}
\def\farcs{\hbox{$.\!\!^{\prime\prime}$}}

\title[Modeling High-Mass Star Formation] 
{Modeling High-Mass Star Formation and Ultracompact \hii\ Regions}

\author[Klessen, Peters, Banerjee, Mac~Low, Keto, Galv{\'a}n-Madrid]   
{Ralf S.\ Klessen$^1$, Thomas Peters$^1$, Robi Banerjee$^1$, 
        Mordecai-Mark Mac Low$^2$, Roberto Galv{\'a}n-Madrid$^{3,4}$ \&
        Eric R. Keto$^3$}

\affiliation{ $^1$Zentrum f\"{u}r Astronomie der Universit\"{a}t Heidelberg,  
 	Institut f\"{u}r Theoretische Astrophysik, Albert-Ueberle-Str. 2, D-69120 Heidelberg, Germany\\[\affilskip] 
 $^2$Department of Astrophysics, American Museum of Natural History,
       79th Street at Central Park West, New York, New York 10024-5192, USA\\[\affilskip]
 $^3$ Harvard-Smithsonian Center for Astrophysics, 60 Garden Street, Cambridge, MA 02138, USA\\[\affilskip]
$^4$ Centro de Radioastronom{\'\i}a y Astrof{\'\i}sica, UNAM, A.P. 3-72 Xangari, Morelia 58089, Mexico}
\pubyear{2010}
\volume{270}  
\pagerange{1 -- 8}
\jname{Numerical Star Formation}
\editors{J.\ Alves, B.\ G.\ Elmegreen, J.\ M.\  Girart \& V.\ L.\ Trimble, eds.}
\begin{document}

\maketitle

\begin{abstract}
 Massive stars influence the surrounding universe far out of proportion to their
  numbers through ionizing radiation, supernova explosions, and heavy element
  production. Their formation requires the collapse of massive interstellar
  gas clouds with very high accretion rates. We discuss results from the first three-dimensional simulations of the gravitational collapse of a massive, rotating   molecular cloud core that include heating by both non-ionizing and ionizing radiation. Local gravitational
  instabilities in the accretion flow lead to the build-up of a small cluster of stars. 
  These lower-mass companions subsequently compete with the high-mass star for the same common gas
  reservoir and limit its overall mass growth. This process is called fragmentation-induced starvation, and explains why massive stars are usually found as members of high-order  stellar systems. These simulations also show that the \hii\ regions forming around massive stars are initially trapped by the infalling gas, but soon begin to fluctuate rapidly.
  Over time, the same
  ultracompact \hii\ region can expand anisotropically, contract again, and take on any of
  the observed morphological classes. The total lifetime of \hii\
  regions is given by the global accretion timescale, rather than their short internal
  sound-crossing time. This solves the so-called lifetime problem of ultracompact \hii\ region.
  We conclude that the the most significant
  differences between the formation of low-mass and high-mass stars are all explained as the
  result of rapid accretion within a dense, gravitationally unstable flow.
\end{abstract}

\firstsection 
\section{Introduction}

High-mass stars form in denser and more massive cloud cores  \citep{motteetal08}
than their low-mass counterparts \citep{myersetal86}. High densities result in
the large accretion rates, exceeding $10^{-4}$~M$_{\odot}$~yr$^{-1}$, required for
massive stars to reach their final mass before exhausting their nuclear fuel \citep{ketoetal06}.
High densities also result in local gravitational instabilities in the accretion flow,
resulting in the formation of multiple additional stars \citep{klesbur00,klesbur01,klessen01,krattmatz06}.
Young massive stars almost always have companions \citep{hohaschik81}, and the number
of their companions significantly exceeds those of low-mass stars \citep{zinnyork07}.
Such companions influence subsequent accretion onto the initial star \citep{krumholzetal09}.
Observations show an upper mass limit of about $100\,$M$_{\odot}$. It remains unclear
whether limits on internal stability or termination of accretion by stellar feedback
determines the value of the upper mass limit \citep{zinnyork07}.

\pagebreak
\hii\ regions form around accreting protostars once they exceed $\sim 10\,$M$_\odot$, equivalent to a spectral type of early B. 
Thus, accretion and ionization must occur
together in the formation of massive stars. The pressure of the 10$^4$~K ionized gas far
exceeds that in the 10$^2$~K accreting molecular gas, creating unique feedback effects such as
ionized outflows \citep{keto02,keto03,keto07}.

Around the most luminous stars the outward radiation pressure force
can equal the inward gravitational attraction. A spherically symmetric
calculation of radiation pressure on dust yields equality at just
under 10 M$_\odot$ \citep{wolfcas87}. However, the dust opacity is
wavelength dependent, the accretion is non-spherical, the
mass-luminosity ratio is different for multiple companions than for a
single star, and to stop accretion more than static force balance is
required. The momentum of any part of the accretion flow must be
reversed \citep{larsstarr71,kahn74,yorkekruegel77,nakanoetal95}. Observations by \citet{ketoetal06}
provide evidence for the presence of all these mitigating
factors, and numerical experiments combining some of these
effects \citep{yorke02,krumkleinmckee07} confirm their effectiveness, showing
that radiation pressure is not dynamically significant below the Eddington limit.

The most significant differences between massive star formation and
low-mass star formation seem to be the clustered nature of star
formation in dense accretion flows and the ionization of these flows.
We present results from three-dimensional simulations by  \citet{petersetal10a,petersetal10b,petersetal10c} of the collapse
of molecular cloud cores to form a cluster of massive stars
that include ionization feedback. These calculations  are the first ones  that allow us to study these effects simultaneously.

\section{Modeling High-Mass Star Formation}

Our discussion is based on a series of recent 
numerical simulations by \cite{petersetal10a,petersetal10b,petersetal10c}. They are
based on a modified version of the adaptive-mesh code
FLASH \citep{fryxell00} that has been extended to include sink particles
representing protostars \citep{federrathetal10}. The protostars evolve following a prestellar
model that determines their stellar and accretion luminosities as function of
protostellar mass and accretion rate. We set the stellar luminosity with
the zero-age main sequence model by \citet{paxton04} and the accretion luminosity by using the tables  by \citet{hosoomu08}.
The ionizing and non-ionizing radiation from the protostars is propagated
through the gas using an improved version of the hybrid characteristics raytracing
method on the adaptive mesh developed by \citet{rijk06}. 
In some calculations, secondary sink formation is suppressed with a density-dependent
temperature floor to prevent runaway collapse of dense blobs of gas. For further details, consult \citet{petersetal10a}.

The simulations start with a $1000\,M_\odot$ molecular cloud. The cloud has a constant density core of $\rho = 1.27 \times 10^{-20}\,$ g\,cm$^{-3}$ within a radius of $r = 0.5\,$pc and then falls off as $r^{-3 / 2}$ until $r = 1.6\,$pc. The initial temperature of the cloud is $T = 30\,$K. The whole cloud is set up in solid body rotation with an angular velocity $\omega = 1.5 \times 10^{-14}\,$s$^{-1}$ corresponding to a ratio of rotational to gravitational energy $\beta = 0.05$ and a mean specific angular momentum of $j = 1.27 \times 10^{23}\,$cm$^2$s$^{-1}$. \citet{petersetal10a,petersetal10b,petersetal10c} follow the gravitational collapse of the molecular cloud with the adaptive mesh until they reach a cell size of $98\,$AU. Then sink particles are created at a cut-off density of $\rho_\mathrm{crit} = 7 \times 10^{-16}\,$g\,cm$^{-3}$. All gas within the accretion radius of $r_\mathrm{sink} = 590\,$AU  above $\rho_\mathrm{crit}$ is accreted to the sink particle if it is gravitationally bound to it. The Jeans mass on the highest refinement level is $M_\mathrm{jeans} = 0.13$\,M$_\odot$. A summary of the three simulations discussed in this proceedings article is provided in Table \ref{tab:colsim}.

\begin{centering}
\begin{table}
\caption{\hspace{0.3cm}Overview of collapse simulations. \label{tab:colsim}}
\hspace*{0.3cm}
\begin{tabular}[t]{ccccccc}
\hline
Name & Resolution & Radiative Feedback & Multiple Sinks &
M$_{\rm sinks}$\,(M$_\odot$) & $N_{\rm sinks}$ & M$_{\rm max}$\,(M$_\odot$) \\
\hline
Run A & 98~AU & yes & no  & 72.13  &  1 & 72.13 \\
Run B & 98~AU & yes & yes & 125.56 & 25 & 23.39 \\
Run D & 98~AU & no  & yes & 151.43 & 37 & 14.64 \\
\hline
\end{tabular}
\end{table}
\end{centering}

\section{Fragmentation-Induced Starvation}

\begin{figure}
\centerline{\includegraphics[width=0.40\textwidth]{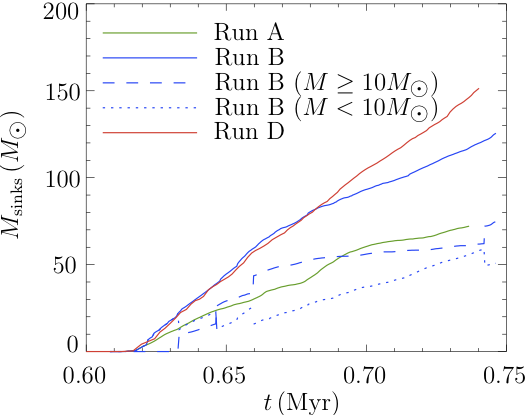}
\includegraphics[width=0.40\textwidth]{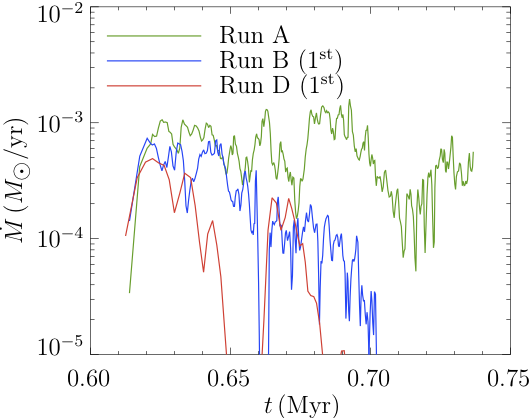}}
\caption{ {\em left} 
Total accretion history of all sink particles combined forming in
Runs A, B, and D. 
{\em right} Instantaneous accretion rate as function of time of the first sink
particle to form in the three runs. 
}
\label{fig:accretion}
\end{figure}

In this section we compare the protostellar mass growth rates from
   three different calculations. Run~A only allows for the formation of a single sink particle, Run~B has multiple sinks
   and radiative heating, while Run~D has multiple sinks but no radiative
   heating.
As already discussed by \citet{petersetal10a},
      when
only the central sink particle
   is
allowed to form (Run~A), nothing
stops the accretion flow to the center. Figure~\ref{fig:accretion} shows that the central
protostar grows at a rate $\dot{M} \approx 5.9 \times 10^{-4}\,$M$_{\odot}\,$yr$^{-1}$ until
we stop the calculation when the star has reached $72\,$M$_{\odot}$. The growing star ionizes
the surrounding gas, raising it to high pressure. However this hot bubble soon breaks out above
and below the disk plane,
   without affecting
the gas flow in the
         disk
midplane much. In
particular, it cannot halt the accretion onto the central star. Similar findings have also been
reported from simulations focussing on the effects of non-ionizing radiation acting on
smaller scales \citep{yorke02,krumkleinmckee07,krumholzetal09,sigalottietal09}.
Radiation pressure cannot stop accretion onto massive stars and is dynamically unimportant,
except maybe in the centers of dense star clusters near the Galactic center.

The situation is different when
         the disk can
fragment and form multiple sink particles.
Initially the mass growth of the central protostar in Runs~B and D is comparable to the one in
Run A. However, as soon as further protostars form in the gravitationally unstable disk, they
begin to compete with the central object for accretion of disk material. Unlike in the
classical competitive accretion picture \citep{bonnell01a,bonetal04}, it is not the most massive
object that dominates and grows disproportionately fast. On the contrary, it is the successive
formation of a number of low-mass objects in the disk at increasing radii that limits subsequent
growth of the more massive objects in the inner disk. Material that moves inwards through the disk
due to viscous and gravitational torques accretes preferentially onto the sinks at larger radii.

Similar behavior is found in models of 
low-mass 
 protobinary disks, where again the secondary accretes at a higher rate than the primary. Its orbit
around the common center of gravity scans larger radii and hence it encounters material that moves
inwards through the disk before the primary star. This drives the system towards equal masses and
circular orbits \citep{bate00}.
In our
simulations, after a certain transition period hardly any gas makes it all the way to the center and
the accretion rate of the first sink particle drops to almost zero. This is the essence of the
fragmentation-induced starvation process. In Run~B, it prevents any
star
         from
reaching a mass
larger than $25\,$M$_\odot$. The Jeans mass in Run~D is smaller than in Run~B because of the
lack of accretion heating, and consequently the highest mass star in Run~D grows to less than
$15\,$M$_\odot$.

Inspection of Figure~\ref{fig:accretion} reveals additional aspects of the process. We see that
the total mass of the sink particle system increases at a faster rate in the multiple sink
simulations, Runs~B and D, than in the single sink case, Run A. This is understandable, because
as more and more gas falls onto the disk it
         becomes
more and more unstable
         to fragmentation, so
as time goes by additional sink particles form at larger
and larger radii.
Star formation occurs
in a larger volume
         of the
disk, and mass growth is not limited by the disk's ability to
transport matter to
         its center by gravitational or viscous
torques.
As a result the overall star-formation
rate is larger than in Run~A.

Since the accretion heating raises the Jeans
    mass and
length in Run~B,
the total number of sink particles is higher in Run~D than in B,
      and the
stars in Run~D generally reach a lower mass than in
Run~B. These two effects
cancel out
to lead to the same overall star formation
          rate for some time. Eventually, however, 
     the total accretion rate of Run~B drops below that of Run~D.
At time $t \approx 0.68\,$Myr
the accretion flow around the most massive star has attenuated below
the value required to trap the \hii\ region. It is able to break out
and affect a significant fraction of the disk area. A comparison with the
mass growth of Run~D clearly shows that there is still enough gas available
to continue constant cluster growth for another $50\,$kyr
or longer, but the gas
     can no longer collapse in Run~B.
Instead, it is swept up in a shell
surrounding the expanding \hii\ region. The figure furthermore demonstrates that,
although the accretion rates of the most massive stars ($M \geq 10\,$M$_\odot$) steadily decrease,
the low-mass stars ($M < 10\,$M$_\odot$), which do not produce any
significant \hii\
         regions,
keep accreting at the same rate.

\section{Properties of the  \hii\ Regions}

In all calculations by \citet{petersetal10a,petersetal10c}, the \hii\ regions are gravitationally trapped in the disk plane but
drive a bipolar outflow perpendicular to the disk.  The highly variable
rate of accretion onto protostars as they pass through dense filaments
causes fast ionization and recombination of large parts of the
interior of the perpendicular outflow. The \hii\ regions around the
massive protostars do not uniformly expand, but instead rapidly
fluctuate in size, shape and luminosity.  

\begin{figure}
\centerline{\includegraphics[width=0.8\textwidth]{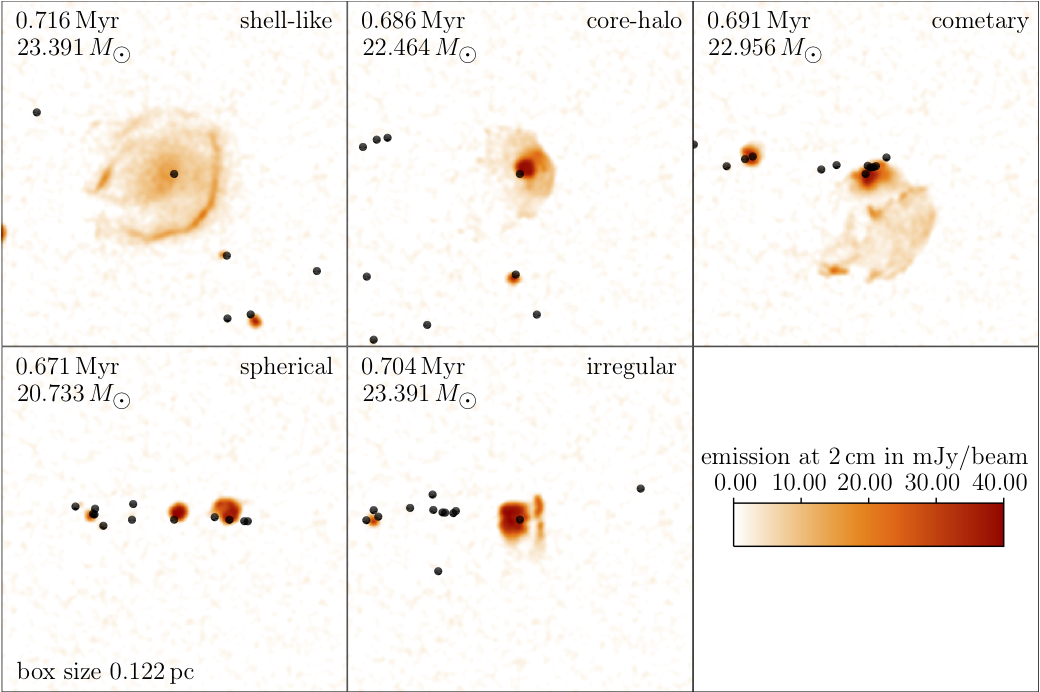}}
\caption{\hii\ region morphologies. This figure shows ultracompact \hii\ regions around massive protostars in Run B at different
time steps and from different viewpoints. The cluster is assumed to be $2.65\,$kpc away, the full width at half maximum of the beam is
$0\farcs14$ and the noise level is $10^{-3}$Jy. This corresponds to typical VLA parameters at a wavelength of $2\,$cm.
The protostellar mass of the central star which powers the \hii\ region is
given in the images. The \hii\ region morphology is highly variable in time and shape, taking the
form of any observed type \citep{woodchurch89,kurtzetal94} during the cluster evolution.\label{fig:morph}}
\end{figure}

We can directly compare these numerical models with radio observations of free-free continuum, hydrogen recombination lines, and NH$_3(3,3)$ rotational lines by generating synthetic maps \citep{petersetal10a,petersetal10b}. The simulated observations of radio continuum emission reproduce the 
morphologies reported in surveys of ultracompact \hii\ regions \citep{woodchurch89,kurtzetal94}. Figure \ref{fig:morph} shows typical images from Run~B to illustrate this point. It is important to note that even the correct relative numbers of the different morphological types are  obtained. Table \ref{tab:morph} shows the   morphology statistics of \uchii\ regions in the surveys of   \citet{woodchurch89}  and \citet{kurtzetal94} as well as   from a random evolutionary sample from Runs~A and B of 500 images for each simulation. While the statistics of the cluster simulation B agrees with the observational data, this is not the case for Run~A, in which only one massive star  forms. For further discussion, see \citet{petersetal10b}.

\begin{table}
\caption{Percentage Frequency Distribution of Morphologies\label{tab:morph}}
\begin{tabular}[t]{ccccc}
\hline
 Type                  & \citet{woodchurch89}    & \citet{kurtzetal94}     & ~~Run A~~     & ~~Run B~~ \\
\hline
Spherical/Unresolved   & 43  & 55  & 19  & 60 $\pm$ 5 \\
Cometary                     & 20  & 16  &  7  & 10 $\pm$ 5 \\
Core-halo                    & 16  &  9  & 15  &  4 $\pm$ 2 \\
Shell-like                   &  4  &  1  &  3  &  5 $\pm$ 1 \\
Irregular                     & 17  & 19  & 57  & 21 $\pm$ 5 \\
\hline
\end{tabular}
\label{statistics}
\end{table}

The \hii\ regions in the model fluctuate rapidly between different shapes while accretion onto the protostar continues. When the gas reservoir around the two most massive stars is exhausted, their \hii\ regions merge into a compact \hii\ region, the type that generally accompanies observed ultracompact \hii\ regions \citep{kimkoo01}. These results suggest that the lifetime problem of ultracompact \hii\ regions \citep{woodchurch89} is only apparent. Since \hii\ regions embedded in accretion flows are continuously fed, and since they flicker with variations in the flow rate, their size does not depend on their age until late in their lifetimes.

\begin{figure}
\hspace{0.05\textwidth}
\includegraphics[height=0.38\textwidth]{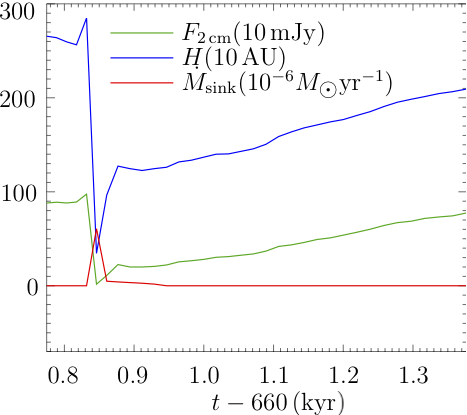}\hspace{0.02\textwidth}
\includegraphics[height=0.38\textwidth]{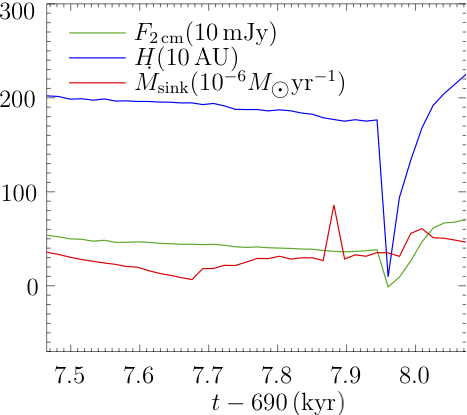}
\caption{\label{fig:time}Time variability of the accretion flow and \hii\ region. The image shows accretion rate onto the sink particle $\dot{M}_\mathrm{sink}$ (sink particle, in units
of $10^{-6}\,$M$_\odot$\,yr$^{-1}$), corresponding diameter $H$ of the \hii\ region (in units of 10 AU), as well as the resulting  2-cm continuum flux $F_{2\mathrm{cm}}$
(in units of 10 mJy) for two selected time periods in Run~B. }
\end{figure}

To compare more directly to observations of the time variability of
	\hii\ regions, \citet{petersetal10b} analyze a few time intervals of interest at a
resolution of 10 years. They find that when the accretion rate
to the star powering the \hii\ region has a large, sudden increase,
the ionized region shrinks, and then slowly re-expands.
This agrees with the contraction, changes in shape, or
anisotropic expansion observed in radio continuum observations of
ultracompact \hii\ regions over intervals of $\sim$
10$\,$yr \citep{francheretal04,rodrigetal07,galvmadetal08}. Figure \ref{fig:time}
shows the 2-cm continuum flux, the characteristic size of the \hii\ region,
and the rate of accretion onto the star. In the left panel, the \hii\ region is initially relatively
large, and accretion is almost shut off. A sudden strong accretion event causes the \hii\ region to shrink and decrease in flux. The star at
this moment has a mass of $19.8\,$M$_\odot$. In the right panel, the star has a larger mass
($23.3\,$M$_\odot$), the \hii\ region is initially smaller, and the star is constantly
accreting gas. The ionizing-photon flux appears to be able to ionize the infalling gas stably,
until a peak in the accretion rate by a factor of 3 and the subsequent continuos accretion of gas
makes the \hii\ region to shrink and decrease in flux. The \hii\ region does not
shrink immediately after the accretion peak because the increase is relatively mild and the
geometry of the infalling gas permitted ionizing photons to escape in one direction. These
results show that observations of large, fast changes in ultracompact \hii\
regions \citep{francheretal04,rodrigetal07,galvmadetal08} are controlled by the accretion process.

\section{Comparison with W51e2}

\begin{figure}
\centerline{\includegraphics[width=0.78\textwidth]{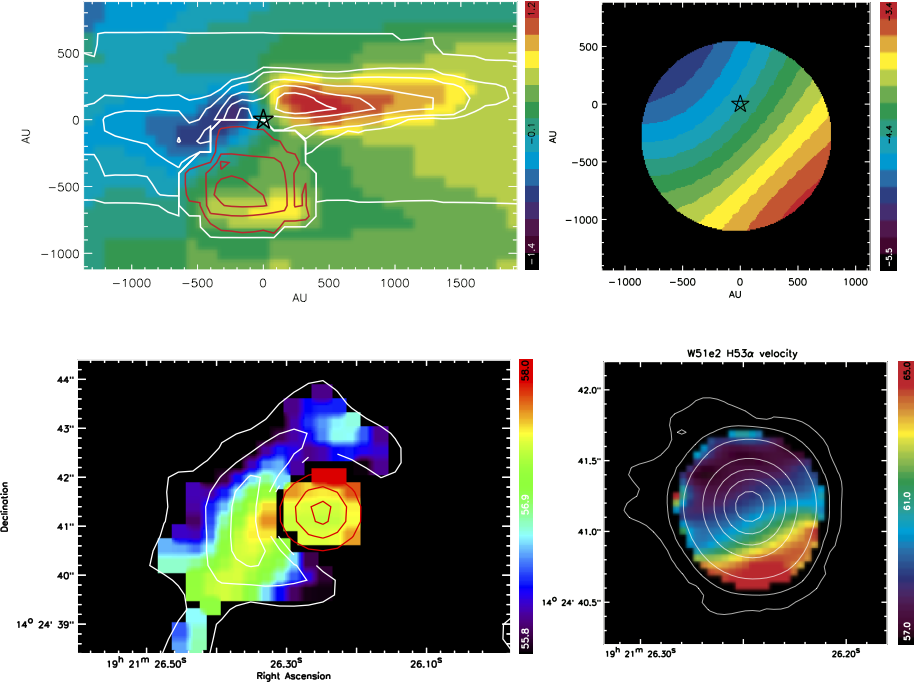}}
\caption{\label{fig:W51e2} Comparison of line and continuum emission simulated from the model (upper panels) and actually observed
from the W51e2 region (lower panels). The left panels show the
NH$_3$(3,3) line emission strength in white contours, the
molecular line velocities as the background color, and the 1.3 cm free-free continuum from ionized
gas in red contours.  The right panels show the H53$\alpha$ recombination line velocities from the ionized gas. The molecular line observations are from
\citet{zhangetal98} and the H53$\alpha$ observations from \citet{ketoklaas08}.
For further details see \citet{petersetal10a}.}
\end{figure}

The \citet{petersetal10a} model can be compared with the well-studied ultracompact \hii\ region
W51e2 \citep{zhangetal98,ketoklaas08}.  In Figure \ref{fig:W51e2} we show
simulated and observed maps of the NH$_3(3,3)$ emission, the 1.3$\,$cm thermal
continuum emission, and the H53$\alpha$ radio recombination line.  The
simulated maps were made at a time when the first star in Run~B has reached a mass of 20$\,$M$_\odot$.

The brightest NH$_3(3,3)$ emission reveals the dense accretion disk
surrounding the most massive star in the model, one of several within
the larger-scale rotationally flattened flow.  The disk shows the
signature of rotation, a gradient from redshifted to blueshifted
velocities across the star. A rotating accretion flow is identifiable in the
observations, oriented from the SE (red velocities) to the NW (blue
velocities) at a projection angle of 135$^\circ$ east of north
(counterclockwise).

The 1.3 cm radio continuum traces the ionized gas, which in the model
expands       perpendicularly to
the accretion disk down the steepest
density gradient. As a result, the simulated map shows the brightest
radio continuum emission just off the mid-plane of the accretion disk,
offset from the central star rather than surrounding it
spherically. Continuum emission in the observations is indeed offset
from the accretion disk traced in ammonia. The NH$_3$(3,3) in front of
the \hii\ region is seen in absorption and red-shifted by its inward
flow toward the protostar. The density gradient in the ionized flow
determines the apparent size of the \hii\ region. Therefore the
accretion time scale determines the age of the \hii\ region rather
than the much shorter sound-crossing time.

Photoevaporation of the actively accreting disk supplies the ionized
flow. Therefore, the ionized gas rotates as it flows outward,
tracing a spiral. An observation that only partially resolves the spatial
structure of the ionized flow sees a velocity gradient oriented in a
direction between that of rotation and of outflow, as shown in the
simulated observation. The observed H53$\alpha$ recombination
line \citep{ketoklaas08} in Figure~4 indeed shows a velocity
gradient oriented between the directions of rotation and the outflow.

\section{Conclusions}
Numerical simulations of high-mass star formation regions are now able to resolve the collapse of massive molecular cloud cores and the accretion flow onto the central group of protostars while at the same time treating the transport of  ionizing and non-ionizing radiation. This allows us to consistently follow the dynamical evolution of the \hii\ regions that ubiquitously accompany the birth of massive stars. We find that the accretion flow becomes gravitationally unstable and fragments. Secondary star formation sets in and consumes material that would otherwise be accreted by the massive  star in the center. We call this process fragmentation-induced starvation. It determines the upper mass limit of the stars in the system and explains why massive stars are usually found as members of larger clusters. These simulations furthermore show that the \hii\ regions forming around massive stars are initially trapped by the infalling gas. But soon, they begin to fluctuate. Over time, the same
  ultracompact \hii\ region can expand anisotropically, contract again, and take on any of
  the observed morphological classes. The total lifetime of \hii\ regions thus is given by the global accretion timescale, rather than their short internal
  sound-crossing time. This solves the so-called lifetime problem of ultracompact \hii\ regions.

\end{document}